\documentclass[a4paper,aps,prd,twocolumn,floats]{revtex4}
\usepackage{graphicx}
\usepackage{amsmath,amssymb,amsfonts}
\usepackage[dvips]{hyperref}

\begin{document}

\title{Testing astrophysical models for the PAMELA positron excess
  with cosmic ray nuclei} 
\author{Philipp Mertsch}
\affiliation{Rudolf Peierls Centre for Theoretical Physics, University
  of Oxford, Oxford OX1 3NP, UK\\\vspace{-0.1ex}}

\author{Subir Sarkar}
\affiliation{Rudolf Peierls Centre for Theoretical Physics, University
  of Oxford, Oxford OX1 3NP, UK\\\vspace{-0.1ex}}

\begin{abstract}
  The excess in the positron fraction reported by the PAMELA
  collaboration has been interpreted as due to annihilation or decay
  of dark matter in the Galaxy. More prosaically, it has been ascribed
  to direct production of positrons by nearby pulsars, or due to pion
  production during stochastic acceleration of hadronic cosmic rays in
  nearby sources. We point out that measurements of secondary nuclei
  produced by cosmic ray spallation can discriminate between these
  possibilities. New data on the titanium-to-iron ratio from the
  ATIC-2 experiment support the hadronic source model above and enable
  a prediction to be made for the boron-to-carbon ratio at energies
  above 100 GeV. Presently, all cosmic ray data are consistent with
  the positron excess being astrophysical in origin.
\end{abstract}

\maketitle


The PAMELA collaboration \cite{Adriani:2008zr} has reported an excess
in the cosmic ray positron fraction, i.e. the ratio of the flux of
positrons to the combined flux of positrons and electrons,
$\phi_{e^+}/(\phi_{e^+} + \phi_{e^-})$, which is significantly above
the background expected from production of positrons and electrons
during propagation of cosmic ray protons and nuclei in the Galaxy
\cite{Moskalenko:1997gh}. It has been noted that the observed rise in
the positron fraction between $\sim 5-100$~GeV cannot be due to
propagation effects \cite{Serpico:2008te}, rather it requires a local
primary source of cosmic ray electrons and positrons, e.g.  nearby
pulsars
\cite{Atoian:1995ux,Hooper:2008kg,Yuksel:2008rf,Profumo:2008ms}. More
excitingly, this could be the long sought for signature of the
annihilation or decay of dark matter particles in the Galaxy
\cite{Bergstrom:2008gr,Cirelli:2008jk,Barger:2008su,Cholis:2008hb,Nardi:2008ix,Arvanitaki:2009yb}.

Alternatively the observed rise in the positron fraction could be due
to the acceleration of positrons produced by the decay of charged
pions, which are created through hadronic interactions of cosmic ray
protons undergoing acceleration in a nearby source
\cite{Blasi:2009hv}. That the secondary-to-primary ratio should {\em
  increase} with energy if secondaries are accelerated in the same
spatial region as the primaries has been noted quite some time ago in
the context of cosmic ray acceleration in the interstellar medium
\cite{Eichler:1980hw, Cowsik:1980ApJ, Fransson:1980Apj}. This model is
conservative since it invokes only processes that are expected to
occur in candidate cosmic ray sources, in particular supernova
remnants (SNR). One way to distinguish it from the other models is to
e.g. compute the expected anti-proton-to-proton ratio, which is
experimentally observed to be {\em consistent} with the standard
background \cite{Adriani:2008zq}. This is in fact in accord with the
above model which predicts a rise in the $\bar{p}$ fraction only at
energies above $\sim 100$~GeV \cite{Blasi:2009bd} (see also
ref.\cite{Fujita:2009wk}). This prediction cannot however be tested
presently but must await data from the forthcoming AMS-2 mission
\cite{AMS02} as well as PAMELA.

Although dark matter annihilation or decay as the explanation of the
positron signal would appear to be disfavoured by the absence of a
corresponding antiproton signal, this can in principle be accomodated
in models with large dark matter particle masses or preferential
leptonic annihilation/decay modes
\cite{Cirelli:2008pk,Cholis:2008qq,Yin:2008bs,Fox:2008kb}. Nearby
pulsars as the source of the positrons are of course quite consistent
with the absence of antiprotons. To differentiate between these
possibilities and the model \cite{Blasi:2009hv} in which secondary
positrons from {\em hadronic} interactions are accelerated in the same
region, we consider secondary nuclei in cosmic rays which are produced
by the spallation of the primaries. An increasing secondary-to-primary
ratio (e.g. boron-to-carbon or titanium-to-iron) in the {\em same}
energy region would confirm that there is indeed a nearby cosmic ray
source where nuclei are being accelerated stochastically along with
protons.

An issue with this model \cite{Blasi:2009hv} is that a crucial
parameter is not known \emph{a priori} but needs to be obtained from
observations. This is the diffusion coefficient of relativistic
particles near the accelerating SNR shock which determines the
importance of a flatter spectral component over the usual Fermi
spectrum and leads to the rise in secondary-to-primary ratios.  Its
absolute value cannot presently be reliably calculated. Observations
of SNR indicate that the magnetic field is quite turbulent so that
relativistic electrons diffuse close to the `Bohm limit' with
diffusion co-efficient: $D^\text{Bohm} = r_\ell c / 3$, where the
Larmor radius $r_\ell$ of the nucleus is proportional to the rigidity
$E/Z$ \cite{Stage:2006xf}. We need to determine the actual diffusion
co-efficient of ions in SNR in ratio to the Bohm value by fitting to
data. A measurement of one nuclear secondary-to-primary ratio
therefore allows us to make predictions for other ratios in the
framework of this model.

Very recently, data on the titanium-to-iron ratio (Ti/Fe) from the
ATIC-2 experiment have been announced \cite{Zatsepin:2009zr} that
indeed show a rise above $\sim 100\, \text{GeV}$. We use this data as
a calibration to determine the diffusion coefficient and then,
extrapolating it according to its rigidity dependence, we predict the
boron-to-carbon ratio (B/C) that should soon be measured by PAMELA
(P.~Picozza, private communication).


Galactic cosmic rays with energies up to the `knee' in the spectrum at
$\sim 3 \times 10^{15} \, \text{eV}$ are believed to be accelerated by
SNRs. The strong shocks present in these environments allow for
efficient diffusive shock acceleration (DSA) by the 1st-order Fermi
process \cite{Blandford:1987pw}. In the simple test-particle
approximation, which is adequate for the level of accuracy of the
present discussion \cite{Malkov:2001}, protons and nuclei that are
injected upstream are accelerated to form a non-thermal power-law
spectrum whose index depends only on the parameters of the shock
front, in particular the compression ratio $r$. For a supersonic shock
with $r=4$ the steady state energy spectrum for protons and nuclei is
$N \mathrm{d}E \propto E^{-\gamma+2} \mathrm{d}E$ where
$\gamma=3r/(r-1)$. In the standard model of galactic cosmic ray
origin, the accelerated primary nuclei produce secondaries by
spallation on hydrogen and helium nuclei in the interstellar medium
(ISM). In the simple `leaky box model' \cite{Cowsik:1967PRD} an
energy-dependent escape of the cosmic rays out of the galaxy is
invoked to obtain a secondary-to-primary ratio that decreases with
energy as observed to date in the region $\sim1-100$~GeV. This is also
obtained by using the GALPROP code \cite{Moskalenko:1997gh} which
solves the full transport equation in 3 dimensions, and can yield both
the time-independent as well as equilibrium solution.

However, as the acceleration time for the highest energy particles is
of the same order as the timescale for spallation, the production of
secondaries inside the sources must be taken into account. In any {\em
  stochastic} acceleration process one then expects the
secondary-to-primary ratio to increase with energy since particles
with higher energy have spent more time in the acceleration region and
have therefore produced more secondaries \cite{Eichler:1980hw,
  Cowsik:1980ApJ, Fransson:1980Apj}. This general argument can be
quantified for the case of DSA by including the production of
secondaries due to spallation and decay as a source term,
\begin{equation}
\!\!\!\! Q_i (\varepsilon_\text{k}) \text{d}\varepsilon_\text{k} 
= \sum_j N_j(\varepsilon_k) 
\big[\sigma^\text{spall}_{j \rightarrow i} \beta c n_\text{gas} +
\frac{1}{\varepsilon_k \tau^\text{dec}_{j \rightarrow i}}\big] 
\,\text{d}\varepsilon_k ,
\end{equation}
where $\varepsilon_k$ is the K.E./nucleon (in GeV), and a loss term, 
\begin{equation}
\Gamma_i \, N_i(\varepsilon_\text{k})
\text{d}\varepsilon_k = N_i(\varepsilon_\text{k}) 
\big[\sigma^\text{spall}_i \beta c n_\text{gas} 
+ \frac{1}{\varepsilon_k \tau^\text{dec}_i}\big] 
\text{d}\varepsilon_\text{k} , 
\end{equation}
where $\sigma^\text{spall}_{j \rightarrow i}$
($\sigma^\text{spall}_i$) and $\tau^\text{dec}_{j \rightarrow i}$
($\tau^\text{dec}_i$) are the partial (total) cross-sections and decay
time, respectively. The transport equation for any nuclear species $i$
then reads
\begin{equation}
  u \frac{\partial f_i}{\partial x} = 
  D_i \frac{\partial^2 f_i}{\partial x^2} 
  + \frac{1}{3} \frac{\mathrm{d}u}{\mathrm{d}x} p \frac{\partial f_i}{\partial p} 
  - \Gamma_i f_i + q_i ,
\label{eqn:Transport}
\end{equation}
where $f_i$ is the phase space density and the different terms from
left to right describe convection, spatial diffusion, adiabatic energy
losses as well as losses and injection of particles from spallation or
decay. We consider the acceleration of {\em all} species in the usual
setup: in the frame of the shock front the plasma upstream ($x<0$) and
downstream ($x>0$) is moving with velocity $u_{-}$ and $u_{+}$
respectively. We solve eq.(\ref{eqn:Transport}) analytically for
relativistic energies $\varepsilon_k$ greater than a few
$\text{GeV/nucleon}$ such that $p \approx E$, $\beta \approx 1$ and
$N_i \mathrm{d}E \approx 4 \pi p^2 f_i \mathrm{d}p$. At these energies
ionization losses can be neglected and the spallation cross sections
become energy independent.

There are three relevant timescales in the problem:
\begin{enumerate}
\item Acceleration time $\tau_{\text{acc}}$:
\begin{align}
  \tau_\text{acc} &= \frac{3}{u_- - u_+} \int_0^p \left( \frac{D_i^+}{u_+} 
  + \frac{D_i^-}{u_-} \right) 
  \frac{\text{d}p'}{p'} \\
  &\simeq 8.8 \, E_{\text{GeV}} Z^{-1} B_{\mu \text{G}} \, \text{yr} \nonumber
\end{align}
for Bohm diffusion and the parameter values mentioned later.
\item Spallation and decay time $\tau_i$.
\begin{equation}
 \tau^\text{spall}_i \equiv 1/\Gamma^\text{spall}_i \sim 1.2 \times 10^7 \, 
\left( \frac{n_{\text{gas}}}{\text{cm}^{-3}} \right)^{-1} \, \text{yr} .
\end{equation}
where an average $\sigma_i$ of ${\cal O}(100)$~mb has been
assumed. The rest lifetime $\tau^\text{dec}_i$ of the isotopes
considered ranges between $4 \times 10^{-2}$~yr and $10^{17}$~yr.
\item Age of the SNR under consideration \cite{Blasi:2009hv,Blasi:2009bd}
\begin{equation}
\tau_\text{SNR} = x_\text{max}/u_+ \sim 2 \times 10^4 \, \text{yr} .
\end{equation}
\end{enumerate}

There are two essential requirements for SNR to efficiently accelerate
nuclei by the DSA mechanism:
\begin{trivlist}
\item (a) $\tau_{\text{acc}} \ll \tau^{\text{spall}}_{i}$, which is
  equivalent to
\begin{equation}
  20 \frac{\Gamma_i^{-} D_i}{u_{-}^2} \ll 1 \quad 
 \Rightarrow \quad \epsilon_{\text{k}} \ll 6.4 \times 10^5 \frac{Z_i}{A_i} \, 
 B_{\mu \text{G}} \, \text{GeV} \, .
\label{eqn:cond1}
\end{equation}
\item (b) $\tau_{\text{SNR}} \ll \tau_i$ which implies,
\begin{equation}
  \frac{x_{\text{max}}}{u_{+}} \ll \frac{1}{\Gamma_i} \quad 
\Rightarrow \quad x \frac{\Gamma_i}{u_{+}} \ll 1\, .
\label{eqn:cond2}
\end{equation}
\end{trivlist}

The isotopes for which condition (b) is not satisfied at the lowest
energy considered viz.  ${}^{56}\text{Ni}$, ${}^{57}\text{Co}$,
${}^{55}\text{Fe}$, ${}^{54}\text{Mn}$, ${}^{51}\text{Cr}$,
${}^{49}\text{V}$, ${}^{44}\text{Ti}$ and ${}^{7}\text{Be}$ do {\em
  not} contribute significantly, so their decays in the source region
are neglected.

We find that the general solution to eq.(\ref{eqn:Transport}) for $x
\neq 0$ is
\begin{eqnarray}
f_i^\pm = \sum_{j \leq i} \left( E_{ji}^\pm \text{e}^{\lambda^\pm_j x/2} 
+ F_{ji}^\pm \text{e}^{\kappa^\pm_j x/2} \right) + G_i^\pm , \\
\text{where}\quad  \lambda_i^\pm = \frac{u_\pm}{D_i^\pm} 
 \big(1 - \sqrt{1 + 4D_i^\pm\Gamma_i^\pm/u^2_\pm}\big) , \nonumber \\
 \kappa_i^\pm   = \frac{u_\pm}{D_i^\pm} 
 \big(1 + \sqrt{1 + 4 D_i^\pm\Gamma_i^\pm/u^2_\pm}\big) ,\nonumber 
\label{eqn:solution}
\end{eqnarray}
where $G_i^\pm$ is the asymptotic value and $E^{+}_{ji}$ and
$F^{-}_{ji}$ are determined by the recursive relations:
\begin{align}
E_{ji}^{\pm} &= \frac{-4 \sum_{m \geq j} E_{mj}^{\pm} 
\Gamma_{j \rightarrow i}^{\pm}}{D_i^{\pm} \lambda_j^{\pm 2} 
- 2 u \lambda_j^{\pm} - 4 \Gamma_i^{\pm}} \, , \label{eqn:Eji} \\
F_{ji}^{\pm} &= \frac{-4 \sum_{m \geq j} F_{mj}^{\pm} 
\Gamma_{j \rightarrow i}^{\pm}}{D_i^{\pm} \kappa_j^{\pm 2} 
- 2 u \kappa_j^{\pm} - 4 \Gamma_i^{\pm}} , 
\label{eqn:Fji}
\end{align}
We require that the phase space distribution function converges to the
adopted primary composition $Y_i$ (at the injection energy $p_0$) far
upstream of the SNR shock:
\begin{equation}
f_i(x,p) \xrightarrow{x \rightarrow -\infty} Y_i \delta (p-p_0) \quad ,
\partial f_i/\partial p \, (x,p) \xrightarrow{x \rightarrow
  -\infty} 0 .
\end{equation}
We also require the solution to remain finite far downstream. As the
phase space density is continuous at the shock front, we connect the
solutions in both half planes to $f_i^0 = f_i(x=0,p)$ and find them to
be:
\begin{align}
f_i^{-} &=  f_i^0 \text{e}^{\kappa^{-}_i x/2} 
+ \sum_{j < i} F_{ji}^{-} \left(\text{e}^{\kappa^{-}_j x/2} 
- \text{e}^{\kappa^{-}_i x/2} \right) \nonumber \\
&+ Y_i \delta (p-p_0) \left( 1 - \text{e}^{\kappa^{-}_i x/2} \right) \, ,
\label{eqn:solution2a} \\
f_i^{+} &= f_i^0 \text{e}^{\lambda^{+}_i x/2} 
+ \sum_{j < i} E_{ji}^{+} \left(\text{e}^{\lambda^{+}_j x/2} 
- \text{e}^{\lambda^{+}_i x/2} \right) \nonumber \\
&+ G_i^{+} \left( 1 - \text{e}^{\lambda^{+}_i x/2} \right) \, .
\label{eqn:solution2b}
\end{align}
Using eqs.(\ref{eqn:cond1}-\ref{eqn:cond2}), we can linearly expand
$\lambda_i^{+}$ and $\kappa_i^{-}$ in eq.(\ref{eqn:solution}) and the
exponentials in eqs.(\ref{eqn:solution2a}-\ref{eqn:solution2b})
\begin{equation}
\text{e}^{\lambda^{+}_i x/2} \simeq 1 - \frac{\Gamma^{+}_i}{u_{+}} \, , \quad
\text{e}^{\kappa^{+}_i x/2} \simeq 
\big(1 + \frac{\Gamma^{-}_i}{u_{-}} \big) \text{e}^{u_{-}x/D_i}  \, 
\end{equation}
 to obtain:
\begin{equation}
f_i^{+} = f_i^0 + \frac{q_i^+(x=0) - \Gamma_i^{+} f_i^0 }{u_{+}} x \, .
\label{eqn:fi+}
\end{equation}
where $q_i^\pm$ denotes the downstream/upstream source term:
$q^{\pm}_i = \sum_{j<i} f_j \Gamma^{\pm}_{i \rightarrow j}$. 

%
%
Finally we integrate the transport equation over an infinitesimal
interval around the shock, assuming that $q_i^{+}/q_i^{-} =
\Gamma_i^{+} / \Gamma_i^{-} = n_{\text{gas}}^{+}/n_{\text{gas}}^{-} =
r$ and that $D_i^{+} \simeq D_i^{-}$:
\begin{align}
&p \frac{\partial f_i}{\partial p} 
= - \gamma f_i^0 - \gamma (1+ r^2) \frac{\Gamma_i^{-} D_i^{-}}{u_{-}^2} 
f_i^0 \nonumber \\
&+ \gamma \left[ (1+ r^2) \frac{q_i^{-}(x=0) D_i^{-} }{u_{-}^2} 
+ Y_i \delta(p-p_0) \right] \, , 
\label{eqn:DEfi0}
\end{align}
which is readily solved by
\begin{align}
&f_i^0(p) =  \int_0^p \frac{\mathrm{d}p'}{p'} 
\left( \frac{p'}{p} \right)^{\gamma} \text{e}^{-\gamma (1 + r^2)  
 (D_i^{-}(p) - D_i^{-}(p')) \Gamma_i^{-}/u_{-}^2} \nonumber \\  
& \!  \! \times  \gamma \left[ (1+ r^2) \frac{q_i^{-}(x=0) D_i^{-} (p')}{u_{-}^2} 
+ Y_i \delta(p'-p_0) \right] .
\label{eqn:fi0} 
\end{align}
Our eqs.(\ref{eqn:fi+}-\ref{eqn:fi0}) should be compared to eqs. (4-6)
of ref.\cite{Blasi:2009hv} where the loss terms $\Gamma_i f_i$ were
not taken into account. The exponential in our eq. (\ref{eqn:fi0})
leads to a natural cut-off in both the primary and secondary spectra
above the energy predicted by eq. (\ref{eqn:cond1}).  However, due to
the approximations we have made, the secondary-to-primary ratios
cannot be predicted reliably for $4 \Gamma_i D_i/u^2 \gtrsim 0.1$
i.e. much beyond $\sim 1$~TeV.

Starting from the heaviest isotope, eqs.(\ref{eqn:fi+}) and
(\ref{eqn:fi0}) can be solved iteratively to obtain the injection
spectrum after integrating over the SNR volume,
\begin{equation}
N_i (E) = 4 \pi \int_0^{u_+ \tau_\text{SN}} \text{d}x\, p^2 f_i (p) \, 4 \pi \, x^2 .
\end{equation}
To account for the subsequent propagation of the nuclei through the
ISM we solve the transport equation in the `leaky box model'
\cite{Cowsik:1967PRD} which reproduces the observed decrease of
secondary-to-primary ratios with energy in the range $\sim1-100$~GeV
by assuming an energy-dependent lifetime for escape from the
Galaxy. The steady state cosmic ray densities $\mathcal{N}_i$ observed
at Earth are then given by recursion, starting from the heaviest
isotope,
\begin{equation}
\mathcal{N}_i = \frac { \sum_{j<i} \left( \Gamma^\text{spall}_{i \rightarrow j} 
+ 1/ \varepsilon_{\text{k}} \tau_{i \rightarrow j} \right) \mathcal{N}_j 
+ \mathcal{R}_\text{SN} N_i }{ 1 / \tau_{\text{esc}, i} + \Gamma_i } \, ,
\end{equation}
where $\mathcal{R}_\text{SN} \sim 0.03$ yr$^{-1}$ is the Galactic
supernova rate.

\begin{figure}[b]
\begin{center}
\includegraphics[width=\linewidth]{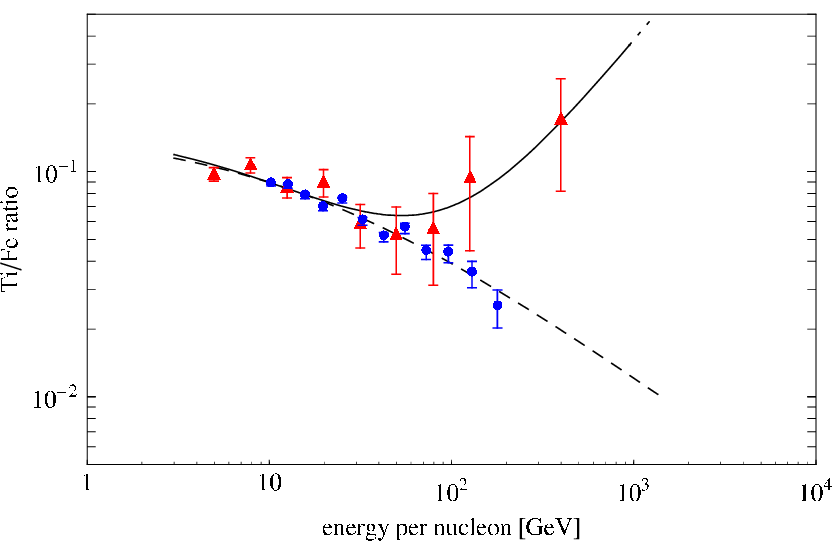}
\end{center}
\vspace{-0.5cm}
\caption[]{The titanium-to-iron ratio in cosmic rays along with model
  predictions --- the `leaky box' model with production of secondaries
  during propagation only (dashed line), and including production and
  acceleration of secondaries in a nearby source (solid line - dotted
  beyond the validity of our calculation). The data points are from
  ATIC-2 (triangles) \cite{Zatsepin:2009zr} and HEAO-3-C3 (circles)
  \cite{Vylet:1990}.}
\label{fig:Ti2Fe}
\end{figure}


We calculate the source densities $N_i$ and ambient densities
$\mathcal{N}_i$, taking into account all stable and metastable
isotopes from ${}^{64} \text{Ni}$ down to ${}^{46} \text{Cr}$/${}^{46}
\text{Ca}$ for the titanium-to-iron ratio, and from ${}^{18} \text{O}$
down to ${}^{10} \text{Be}$ for the boron-to-carbon ratio. Short lived
isotopes that $\beta^{\pm}$ decay immediately into (meta)stable
elements are accounted for in the cross-sections. The primary source
abundances are taken from ref.\cite{Engelmann:1990aa} and we have
adopted an injection energy of 1~GeV independent of the species. The
partial spallation cross-sections are from semi-analytical tabulations
\cite{Silberberg:1973+1977} and the total inelastic cross-sections is
obtained from an empirical formula \cite{Silberberg:1990nj}. The
escape time is modelled according to the usual relation:
\begin{equation}
\tau_{\text{esc}, i} = \rho \, c \, x_{\text{esc}, i} = \rho \, c \,
x_{\text{esc}, i}^0 (E/Z_i)^{-\mu}
\end{equation}
where $x$ is the column density
traversed in the ISM and $\rho = 0.02 \, \text{atom} \,
\text{cm}^{-3}$ is the typical mass density of hydrogen in the
ISM. 
We have neglected spallation on helium at this level of precision as
its inclusion will have an effect $<10 \%$. The parameters of the fit
are sensitive to the adopted partial spallation cross-sections, for
example $\mu \simeq 0.7$ for the Ti/Fe ratio but $\sim 0.6$ for the
B/C ratio. For improved accuracy a current compilation of
experimentally deduced cross-sections should be used.

Following ref.\cite{Blasi:2009bd}, the parameters are chosen to be:
$r=4$, $u_{-} = 0.5 \times 10^8 \, \text{cm} \,\text{s}^{-1}$,
$n_\text{gas}^{-} = 2 \,\text{cm}^{-3}$ and $B = 1 \, \mu
\text{G}$. The diffusion coefficient in the SNR is
\begin{equation}
D_i (E) = 3.3 \times 10^{22} \mathcal{F}^{-1} \, B_{\mu}^{-1} 
E_{\text{GeV}} Z_i^{-1} \, \text{cm}^2 \text{s}^{-1}
\label{eqn:D(p)}
\end{equation}
where the fudge factor $\mathcal{F}^{-1}$ is the ratio of the
diffusion coefficient to the Bohm value and is determined by
fitting to the measured titanium-to-iron ratio.

\begin{figure}[t]
\begin{center}
\includegraphics[width=\linewidth]{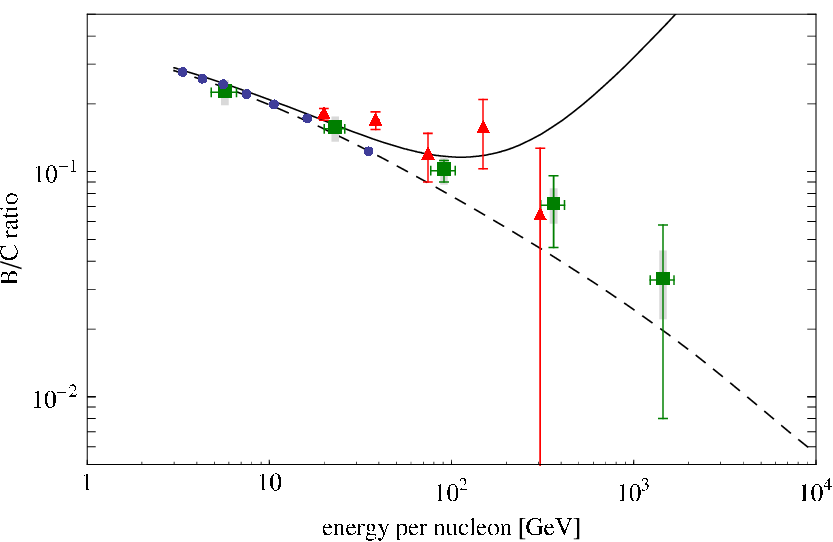}
\end{center}
\vspace{-0.5cm}
\caption[]{ The boron-to-carbon ratio in cosmic rays along with model
  predictions --- the `leaky box' model with production of secondaries
  during propagation only (dashed line), and including production and
  acceleration of secondaries in a nearby source (solid line). The
  data points are from HEAO-3-C2 (circles) \cite{Engelmann:1990aa},
  ATIC-2 (triangles) \cite{Panov:2007fe} and CREAM (squares)
  \cite{Ahn:2008my}.}
\label{fig:B2C}
\end{figure}

The calculated titanium-to-iron ratio together with relevant
experimental data is shown in Fig. \ref{fig:Ti2Fe}. The dashed line
corresponds to the leaky box model with production of secondaries
during propagation only and is a good fit to the (reanalysed)
HEAO-3-C3 data \cite{Vylet:1990}. The solid line includes production
and acceleration of secondaries inside the source regions which
results in an {\em increasing} ratio for energies above $\sim 50 \,
\text{GeV}/n$ and reproduces well the ATIC-2 data
\cite{Zatsepin:2009zr} taking $\mathcal{F}^{-1} \simeq 40$. This is
similar to the value reported in
refs.\cite{Blasi:2009hv,Blasi:2009bd}, thus ensuring consistency with
the $e^+$ as well as $\bar{p}$ fraction measured by PAMELA.

Clearly the experimental situation is inconclusive so a new test is
called for. Fig. \ref{fig:B2C} shows the corresponding expectation for
the boron-to-carbon ratio with the diffusion coefficient scaled
proportional to rigidity according to eq.(\ref{eqn:D(p)}). The CREAM
data \cite{Ahn:2008my} do show a downward trend as has been emphasized
recently \cite{Simet:2009ne}, but the uncertainties are still large so
we await more precise measurements by PAMELA which has been directly
calibrated in a test beam \cite{Campana:2008xj}. Agreement with our
prediction would confirm the astrophysical origin of the positron
excess as proposed in ref.\cite{Blasi:2009hv} and thus establish the
existence of an accelerator of hadronic cosmic rays within a few kpc.

\section*{Acknowledgements}

PM is grateful to Markus Ahlers for many helpful and inspiring
discussions and SS thanks Ramanath Cowsik for an inspiring remark in
1979 as well as Pasquale Blasi and Dan Hooper for helpful comments.
PM is partly funded by a STFC Postgraduate Studentship. Both authors
acknowledge support by the EU Marie Curie Network ``UniverseNet''
(HPRN-CT-2006-035863).

\end{document}